\documentclass[preprint,amsmath,amssymb,onecolumn,showpacs,floatfix,aps,11pt,nofootinbib]{revtex4}
\usepackage{subfigure}

\newcommand{\re}{\mathtt{Re}}
\newcommand{\img}{\mathtt{Im}}
\newcommand{\CCC}{C}

\newcommand{\II}{\mathbb{I}}

\newcommand{\bec}{\begin{center}}
\newcommand{\eec}{\end{center}}

\newcommand{\bea}{\begin{array}}
\newcommand{\ear}{\end{array}}

\newcommand{\bfr}{\begin{flushright}}

\newcommand{\efr}{\end{flushright}}

\newcommand{\RR}{\mathbb{R}}

\newcommand{\bege}{\begin{equation}}
\newcommand{\enge}{\end{equation}}

\newcommand{\g}{\gamma}

\newcommand{\beq}{\begin{eqnarray}}\newcommand{\benu}{\begin{enumerate}}\newcommand{\enu}{\end{enumerate}}
\newcommand{\eeq}{\end{eqnarray}}

\newcommand{\CC}{\mathbb{C}}

\newcommand{\OO}{\mathbb{O}}

\newcommand{\KKKK}{{\bf{\mathcal{K}}}}

\newcommand{\bx}{\begin{pmatrix}}
\newcommand{\ex}{\end{pmatrix}}

\usepackage{pict2e} % melhora o backslashbox
\usepackage{rotating}

\begin{document}

%\preprint{}

\title{Classification of Singular
Spinor Fields and Other Mass Dimension One Fermions}


\author{R. T. Cavalcanti}
\email[]{rogerio.cavalcanti@ufabc.edu.br}
%\thanks{}
\affiliation{Centro de Ci\^encias Naturais e Humanas,
UFABC \\ Santo Andr\'e - SP,09210-580, Brasil}


\date{\today}

\begin{abstract}
We investigate the constraint equations of the Lounesto spinor fields classification and show that it can be used to completely characterize all the singular classes, which can potentially accommodate further mass dimension one fermions, beyond the well known Elko spinor fields. This result can be useful for two purposes: besides a great abridgement
 in the classification of a given spinor field, we provide a general form of each class of spinor fields, which can be used furthermore to search for a general classification of spinors dynamics.

\end{abstract}


\maketitle

\section{Introduction}

Classification and characterization results are useful in all fields of Science. In Physics, specifically,  it can be used to choose the suitable structure to describe phenomena which are still not understood thoroughly. In this sense it would be desirable that the spinor fields classification classes, in Lounesto spinor field classification, could be characterized in the simplest way, and hence be used straightforwardly thereon. In fact it is one of the two main purposes of this paper, introducing Lounesto spinor field classification \cite{lou2} and characterizing its classes. It turns out that the classification process can be simplified, avoiding tedious calculations and  intricate concepts from Clifford algebras that are dispensable in a first moment. In spite of this abridgement,  it could be furthermore helpful in the next step of the spinors fields characterization, namely, by finding all the possible dynamics associated to each one of the spinor classes in Lounesto spinor field classification \cite{unfold, advapp}. We already know that there are   
type-(6) spinor fields satisfying the Weyl equations, spinor fields of types-(1), -(2), and (3) satisfying the
Dirac equation, spinor fields of type-(4) (flag-dipoles) satisfying a Dirac equation in a framework provided by $f(R)$ gravity with torsion \cite{esk}, and spinor fields of type-(5) (flagpoles) governed by Majorana \cite{mako} or  
Elko coupled system of equations \cite{allu, allu1,wa1}, for instance. 
Nevertheless, the above mentioned types of 
dynamics do not cover all the spinor fields in their respective classes, and hence these spinor  fields dynamics seem to be merely 
the tip of the iceberg. Quite different spinor fields with different mass dimension in one of the Lounesto classes --- for example Majorana and Elko spinor fields in class-(5) of Lounesto classification --- 
can present completely contrastive dynamics.
Before departing to solve the general question 
of determining all the possible dynamics, 
we need to first to ascertain which types 
of spinor fields can still exist in the given classes of Lounesto classification, apart from the well known ones mentioned above.

It is the second main purpose of the paper, paving the first step for this ambitious achievement. Some effort have been made in looking for the dynamics of each spinor class under Lounesto classification, see, e. g. Refs. \cite{allu,allu1,crevice} for type-(5) spinor fields and Refs. \cite{unfold,advapp,esk,villalobos,exotic} for other type-(4) ones, thus the algebraic characterization introduced in this work can provide tips toward what direction should be followed.

Classical spinor fields are well known to carry irreducible representations of the Lorentz group,  what imply the usual Dirac, Weyl, and Majorana spinor fields.
 Lounesto spinor field classification suggests new types of spinor fields, as the flag-dipole spinor fields recently found as a particle \cite{vign,esk}. Moreover, another class of spinor fields in this classification known as  flagpoles encompasses Elko spinor fields \cite{wa1},  candidates to the dark matter description \cite{allu, allu1, horv1,bur2,lee1,lee2,aaa,boeh1,sh1,sh} with  applications in field theory and cosmology \cite{alex,marcao,bur1,sadj,liu}. 
Lounesto spinor field classification has 
wide applications, not solely in quantum field theory, and supersymmetry \cite{wund}, but additionally in gravity \cite{boeh,boeh1,bur2}. For instance,  the Einstein-Hilbert, the Einstein-Palatini, and the Holst actions were shown to be derived from the quadratic spinor Lagrangian, when the three classes of Dirac spinor fields, under Lounesto spinor field classification, are considered \cite{ro11,ro12}. Moreover, 
a straightforward mass dimension transmutation path from Dirac to Elko spinor fields have been established in Refs. \cite{jmp07,ij9}.

As Elko spinor fields have mass dimension one \cite{allu, allu1,crevice}, there is nothing that precludes the appearance of mass dimension one spinor fields further in classes (4) and (6) in Lounesto spinor field classification. Although there is no quantum field constructed out with type-(4) spinor fields yet, it does not respect the full Lorentz symmetry \cite{unfold}, being  another potential candidate to the dark matter. Type-(4) spinor fields are wealthy regarding its mathematical structure \cite{lou2,unfold,advapp} and from the Physics point of view, as above mentioned,  they have been found to be a particle corresponding to the solution of the Dirac equation in $f(R)$ gravity with torsion \cite{esk}. Thus at least type-(4) spinor fields 
are also potential candidates to construct mass dimension one fermions.

This paper is organized as follows: in Section II we briefly introduce the bilinear covariants and the Lounesto's algebraic classification of spinor fields. In Section III we discuss Elko spinor field, a mass dimension one fermion which is the main representative of type-(5) class and a prime candidate to describe dark matter. A brief summary of its properties is provided. Finally, in Section IV we show how the constraint equation coming from Lounesto's classification can be used to completely characterize singular spinors. The results are summarized in a  characterization of singular spinor fields table at the end of the section.

\section{Bilinear Covariants and Spinors Classification}

The multivector structure of the space-time Clifford algebra is well known to impose five objects preserving Lorentz symmetry, the so called bilinear covariants \cite{yvon,taka,fierz,hol,cra}. They carry the observable physical information of the Dirac theory.  According to their multivector structure and number of components, the bilinear covariants are listed below:
\begin{eqnarray}
\mbox{Scalar (1)}&:&\sigma = \psi^{\dagger}\g_0\psi\\
\mbox{Vector (4)}&:&J_{\mu} = \psi^{\dagger}\g_0\g_{\mu}\psi \\
\mbox{Bivector (6)}&:&S_{\mu \nu} = i\psi^{\dagger}\g_0\g_{\mu}\g_{ \nu}\psi \\
\mbox{Pseudovector (4)}&:&K_{\mu} = \psi^{\dagger}\g_0\g_{5}\g_{\mu}\psi \\
\mbox{Pseudoscalar (1)}&:&\omega = i\psi^{\dagger}\g_0\g_{5}\psi
\end{eqnarray}
where $J_\mu, S_{\mu\nu}$ and $K_\mu$ are the components of $\mathbf{J}, \mathbf{S}$ and $\mathbf{K}$ respectively \cite{lou2}. Here and in what follows, we have adopted the metric $\eta^{\mu \nu}=(+,-,-,-)$.

%The vector $({J}^0,J^k)$ is time like, future oriented and the component $J^0$ gives the probability density (of what?***).  The components $J^k, \; k=1,2,3$ are associated to the density current (of what?***) and obey the continuity equation $\partial_{\mu}J^{\mu}=0.$
%The bivector $\textbf{S}$ is associated to the intrinsic angular momentum and  ${K}^\mu$ are components of the electron spin direction.% \cite{lou2, ryder, rqm}. 

%\subsection*{Spinor Fields Classification}
Usually spinor fields are classified according to irreducible representations of the spin group $Spin_+(1,3)$ connected to the identity, classically known as Weyl, Dirac, and Majorana. On the other hand, as showed by Lounesto,  another classification can be defined using bilinear covariants \cite{lou2}. Lounesto's results have emerged in the Clifford algebras context, thus it does not depend on any  representation. However, by choosing a representation it turn the classification straightforward to be applied in Physics. %Some theoretical efforts have been made to identify how non standard spinor fields, in a sense that will be clear below, can appear in Nature. 
The classification is given below, where $\mathbf{J}$ is always nonzero and in the first three classes  $\mathbf{K}$\textbf{, }$\mathbf{S}$ $\neq0$: %(***estes ``vetores'' nao foram definidos antes: defina-os)***
\begin{center}
\begin{tabular}{cccccccc}
\hline 
(1) & $\sigma \neq 0, \omega \neq 0$ & \qquad &&&& (4) & $\sigma =0= \omega, \textbf{K} \neq 0, \textbf{S} \neq 0$ \\ 
(2) & $\sigma \neq 0, \omega = 0$ & \qquad &&&& (5) & $\sigma =0= \omega, \textbf{K} = 0, \textbf{S} \neq 0$ \\ 
(3) & $\sigma=0, \omega \neq 0$ & \qquad &&&& (6) & $\sigma =0= \omega, \textbf{K} \neq 0, \textbf{S} = 0$ \\
\hline
\end{tabular} 
\end{center}

\noindent 
Type-(1), (2), and (3)  are called Dirac spinors or regular spinors.  On the other hand  type-(4), (5) e (6) are called singular spinors, or individually \textit{flagpole}, \textit{flag-dipole} and Weyl spinors respectively. Regular and type-(6) spinor fields are well known from quantum field theory, however until recently type-(4) and type-(5) had not received substantial attention in the physics community, thus they could be called non standard.  %It has been changing after non standard spinors were shown to be  physical solutions of the Dirac equation in the context of ESK gravity  \cite{vign,esk}. 

\section{Elko Spinor Fields}

In the following we are going to use the Weyl representation of $\gamma^{\mu}$, given by

\begin{equation}
\gamma^{0}=%
\begin{pmatrix}
\OO_2 & \II_2\\
\II_2 & \OO_2
\end{pmatrix}
,\;\;\gamma^{k}=%
\begin{pmatrix}
\OO_2 & -\sigma_{k}\\
\sigma_{k} & \OO_2
\end{pmatrix}
,\;\;
\gamma^{5}=i\gamma^{0}\gamma^{1}\gamma^{2}\gamma^{3}=\begin{pmatrix}
\II_2&\OO_2 \\
\OO_2 & -\II_2
\end{pmatrix},
\label{dirac matrices}%
\end{equation}
\noindent where
 $\sigma_i$ are the Pauli matrices.

Elko spinors are eigenspinors of the charge conjugation operator\footnote{From the German ``Eigenspinoren des
Ladungskonjugationsoperators''.}. Due to its very small coupling with the standard model fields, except by the Higgs field, it was proposed as a prime candidate to describe dark matter \cite{allu,allu1,crevice,horv1,lee1,lee2}. Some results have been revised after the first papers proposing Elko, including its symmetries \cite{horv1} and Lagrangian \cite{lee3}, but these new results have kept the fundamental 
physical properties of this kind of spinor fields \cite{crevice,lee3}. An updated publication concerning Elko spinor fields can be found in Ref. \cite{crevice}.

The Elko spinor can be expressed in general as   
\begin{equation}
\lambda({k^\mu})=\binom{\sigma_2\phi^{\ast}(k^\mu)}{\phi(k^\mu)}, \qquad k^\mu \equiv \lim_{p\to 0}\left(m,{{\bf p}}\right)
\label{1}%
\end{equation}
\noindent  where $\phi(k^\mu)$ denotes a left-handed Weyl
spinor and $p = \|{\bf p}\|$. Being eigenspinors of the charge
conjugation operator $C$, it holds $C\lambda(k^\mu)=\pm \lambda(k^\mu)$. %, for $
%C=%
%\begin{pmatrix}
%\OO & i\Phi \\
%-i\Phi & \OO
%\end{pmatrix}\,K.$  
%The operator $K$ is responsible for the $\mathbb{C}$-conjugation of
%spinor fields appearing on the right. 
The plus and minus sign regards {self-conjugate} and {anti self-conjugate}  spinor fields, denoted by $\lambda^{S}(k^\mu)$ and $\lambda^{A}(k^\mu)$.

  Explicitly, the complete form of Elko spinor fields can be
found by solving the equation of helicity
$(\sigma\cdot\widehat{\bf{p}})\phi^{\pm}(k^\mu)=\pm
\phi^{\pm}(k^\mu)$ in the rest frame and subsequently
performing a boost \cite{allu, crevice}. 
%Note that the helicity of
%$i\Phi[\phi_{L}(\mathbf{p})]^\ast$ is opposed to that of
%$\phi_L(\mathbf{p})$, since
%$(\sigma\cdot\widehat{\bf{p}})\Phi[\phi_L^{{\pm}}(\mathbf{0})]^\ast=\mp
%\Phi[\phi_L^{\ast\pm}(\mathbf{0})]^\ast$. Here
%$\widehat{\bf{p}}:={\bf p}/|{\bf p}|=(\sin\theta\cos\phi,\sin\theta\sin\phi,\cos\theta)$. 
Therefore Elko 
spinor fields are given by \cite{crevice}
\begin{align}
\lambda^{S}_{\pm}(p^\mu)&=\sqrt{\frac{E+m}{2m}}\Bigg(1\mp
\frac{{p}}{E+m}\Bigg)\lambda^{S}_{\pm}(k^\mu),\\
\lambda^{A}_{\pm }(p^\mu)&=\sqrt{\frac{E+m}{2m}}\Bigg(1\pm
\frac{{p}}{E+m}\Bigg)\lambda^{A}_{\pm }(k^\mu),
\end{align}
where  
\begin{equation}
\lambda^{S}_{\pm}(k^\mu)=%
\begin{pmatrix}
  \sigma_2[\phi^{\pm}(k^\mu)]^{*} \\
\phi^{\pm}(k^\mu)
\end{pmatrix}\;\;\;\mbox{and}\;\;\;\lambda^{A}_{\pm}(k^\mu)=%
\pm\begin{pmatrix}
- \sigma_2[\phi^{\mp}(k^\mu)]^{*} \\
\phi^{\mp}(k^\mu)
\end{pmatrix}\,.
\end{equation}

There are several interesting and unusual aspects concerning Elko theory, including its mass dimension and relationship with the very special relativity (VSR) groups. Actually Ahluwalia and Horvath showed that the Elko theory is invariant under the action of the HOM(2) VSR group and covariant under SIM(2) VSR group\cite{horv1}. %In fact, when the parity operator is absent in a given relativistic theory, it is possible to rebuilt the dynamical objects by thinking of irreducible representations of  subgroups of the Lorentz group \cite{cohen}. 
Without going into specific details, Cohen and Glashow argued that Very Special Relativity, rather than Special Relativity, could be the fundamental symmetry of nature. VSR is realized taking some proper subgroups of the orthochronous proper Lorentz group, specially the so called HOM(2) and SIM(2) subgroups. In fact they showed that the group HOM(2), which is a subgroup of SIM(2), is necessary and sufficient to \cite{cohen}:

\begin{itemize}

\item Explain the results of the Michelson-Morley experiments and its more sensitive results.

\item Ensure that the speed of light is the same for all observers.

\item Preserve SR time dilatation and the law of velocity addition.
\end{itemize}

Furthermore all VSR subgroups share the property that incorporates either P, T, CP, or CT symmetries enlarges these subgroups to the full Lorentz group.

Under Lounesto classification, Elko is a representative of mass dimension one fermion in the type-(5) spinor field class, but is not the most general, as also Majorana spinor fields are encompassed by such class. Thus if one is looking for a true classification of the spinor field dynamics based on Lounesto classes, as previously mentioned, the knowledge of the general form of spinor fields of each class is a critical point. It is exactly the aim of the next section, to characterize each spinor field class under Lounesto classification.

\section{Algebraic Spinor Field Characterization}

This section finally provides the two main purposes of the paper. The first is to simplify the use of Lounesto classification, allowing us to identify the spinor field  class just looking for its components. The second one is to provide  a general form of spinor fields of each class, which can be useful for finding its dynamics and for identifying possible other spinor fields of mass dimension one, among the singular classes. 

We can realize a simplification answering the following question: can the constraint equations that define a class be used to characterize the spinor fields? Fortunately for singular spinor fields the answer is positive, as we are going to see throughout the present section. 
To achieve our goal, first we need  to know all the bilinear covariants components for a general spinor field, given by $\psi(x)=\left(a(x),b(x),c(x),d(x)\right)^T$, where $a,b,c,d:\RR^{1,3} \to \CC$. As a matter of  simplicity, the $x$ dependence in the components are going to be omitted\footnote{For further properties on spinor fields see, e. g., \cite{hidden}.}. By performing some simple but tedious calculation the following components emerge: 
%\begin{footnotesize}

\begin{align}
\sigma = 2\mathtt{Re} [ac^*+bd^*]\;; & \quad \omega = -2\mathtt{Im}[ac^*+bd^*]\\% \nonumber\\
\textbf{J}  =\begin{pmatrix}
 ||a||^2+|| b ||^2 +  || c ||^2+||d||^2\\ 
2\re[ab^*-cd^*] \\ 
2\img[ab^*-cd^*] \\ 
||b||^2+||c||^2-||a||^2-||d||^2
\end{pmatrix} \;; & \quad\textbf{K} =\begin{pmatrix}
||c||^2+||d||^2-||a||^2-||b||^2\\ 
-2\re[ab^*+cd^*]\\ 
-2\img[ab^*+cd^*]\\ 
||a||^2+||c||^2-||b||^2-||d||^2
\end{pmatrix}\\ %\nonumber \\
\textbf{S}  =\left\{\begin{array}{lcr}
S_{01}&=&-2\img[ad^*+bc^*]\;\;; \\ 
S_{03}&=& 2\img[ac^*-bd^*]\; \;;\\ 
 S_{23}&=& 2\re[ad^*+bc^*]\;\;;
\end{array}  \right. & \quad \begin{array}{lcr}
S_{02}&=&2\re[ad^*-bc^*]  \\ 
S_{12}&=&-2\re[ac^*-bd^*] \\ 
S_{31} &=&2\img[ad^*-bc^*]
\end{array} %\\
%S_{01}=2\img[ad^*+bc^*] \;; & \quad S_{02}=2\re[ad^*-bc^*]  \\ 
%S_{03}= 2\img[ac^*-bd^*]\;; & \quad S_{12}=2\re[ac^*-bd^*]\\
% S_{23}= 2\re[ad^*+bc^*] \;; & \quad S_{31} -2\img[ad^*-bc^*]
\end{align}
%\begin{align}
%\begin{array}{ccc}
%S_{01}=2\img[ad^*+bc^*] ;& S_{02}=2\re[ad^*-bc^*] ;&S_{03}= 2\img[ac^*-bd^*] \\ 
%S_{12}=2\re[ac^*-bd^*] ; &S_{23} 2\re[ad^*+bc^*] ;&S_{31} -2\img[ad^*-bc^*]\\
%S_{12}=2\re[ac^*-bd^*] ; &S_{23} 2\re[ad^*+bc^*] ;&S_{31} -2\img[ad^*-bc^*]
%\end{array} 
%\end{align}
%\begin{align*}
%\sigma = &2\mathtt{Re} [ac^*+bd^*],\qquad  \omega = -2\mathtt{Im}[ac^*+bd^*]\\ 
%\textbf{J}  =&\left(||a||^2+|| b ||^2 +  || c ||^2+||d||^2, 2\re[cd^*-ab^*],-2\img[cd^*-ab^*],\right.\\
%&\left.||b||^2+||c||^2-||a||^2-||d||^2\right)\\
%\textbf{K}  =&\left(||c||^2+||d||^2-||a||^2-||b||^2,2\re[cd^*+ab^*], -2\img[cd^*+ab^*],\right.\\
%&\left.||a||^2+||c||^2-||b||^2-||d||^2\right)\\
%\textbf{S}  =&\left(2\img[ad^*+bc^*],2\re[ad^*-bc^*], 2\img[ac^*-bd^*],2\re[ac^*-bd^*], 2\re[ad^*+bc^*],\right.\\
%&\left. -2\img[ad^*-bc^*] \right)
%\end{align*}
%\end{footnotesize}

where we have used $ A=\re[A]+i\img[A]$.

\subsection{Singular and Regular Spinor Fields}

The property shared by every singular spinor field gives us the first tip in the route of characterization, namely 
\begin{equation}\label{eq1}
\sigma = 0 =  \omega \Rightarrow a c^{*}+b d^{*}=0.
\end{equation} 
The above equation  must be true for singular spinor fields in any referential frame. Hence when one is looking for an equation to describe this kind of spinor field, the constraint $ac^{*}=-b d^{*}$ must be preserved. In particular, when all the components are non zero, one of them can be written in terms of the others, as for instance $a=-\frac{bcd^*}{||c||^2}$. In the following we combine the additional constraints to characterize singular spinor fields of each class.

\subsection*{Type-(6)}

The spinor fields belonging to this class encompass the best known singular spinor fields, the so called Weyl spinor fields \cite{lou2}. At first, to study this characterization could sound unnecessary, however as a matter of completeness we  deal with them. In spite of completeness, it shall be useful when we look for the less symmetrical case, the type-(4) class. 

Type-(6) spinor field class implies
\begin{align}
\left.\begin{array}{rrllcl}
\sigma = &0&  \\
 \omega = &0&\\
\mathbf{S} = &0& 
\end{array}\right\} \Rightarrow \left\{ \begin{array}{ccc}
ac^* & = & 0\\
bd^* & = & 0\\
ad^* & = & 0\\
bc^* & = & 0
\end{array} \right.
\end{align}
thus $a=0=b$ or $c=0=d$, and therefore no more than two components can be non zero. The two possibilities, as expected, are
\begin{eqnarray}
\psi_{_{(6)}}^-=\left(a,b,0,0\right)^T \qquad \mbox{ or }\qquad \psi_{_{(6)}}^+=\left(0,0,c,d\right)^T\,.
\end{eqnarray}
The components $K_0$ or $K_3$ ensure $\textbf{K} \neq 0$, even when just one spinor field component is non zero.

\subsection*{Type-(5)}
%***citar mais%Two type-(5) spinors raised in physics research, the so called Majorana and Elko spinor filds. The first one...**(um pouco mais sobre Majorana). Elko, in the other hand, is a prime candidate to describe dark matter \cite{allu, allu1} **citar mais. Despite have some interesting and exotic properties , from the point of view of topological structure, locality, ...**(um pouco mais sobre Elko)
%\cite{exotic,alex}
In addition to the equation shared by all singular spinor fields, for type-(5) class we have three more, provided by: 
\begin{equation}\label{eq3}
\mathbf{K}=0 \Longrightarrow b a^{*}+d c^{*}=0, ||a||^2=||d||^2 \mbox{ and } ||b||^2=||c||^2\,.
\end{equation}  
Therefore it must look like one of the following:
\begin{eqnarray}
\psi_{_{(5)}}=\left(a,0,0,d\right)^T\; ;  \qquad \psi_{_{(5)}}=\left(0,b,c,0\right)^T\; ;  \qquad  \psi_{_{(5)}}=\left(-\frac{bcd^*}{||c||^2},b,c,d\right)^T.
\end{eqnarray}
By combining eq. \eqref{eq3} and eq. \eqref{eq1} we can improve the previous result as follows. From the expressions 
\begin{equation}\label{eq4}
a=-d^*(c + b)(c^*+b^*)^{-1}=-d^*\left[\frac{c+b}{\Vert c+b \Vert}\right]^2
\end{equation} 
and 
\begin{align}
\tan\varphi_1=-i\frac{c+b-(c+b)^*}{c+b+(c+b)^*}
\end{align}
we have $a=-d^*e^{2i \varphi_1}$ and $b=c^*e^{2i \varphi_1}$, so type-(5) spinor fields can be settled in a more representative way\footnote{If $a \neq -d$ or $b \neq -c$, otherwise would be type-(2).} by $\psi_{_{(5)}}=\left( -d^*e^{2i \varphi_1},c^*e^{2i \varphi_1},c,d\right)^T.$ 
%\begin{equation}\label{eq5}
%a=-d^*e^{2i \varphi_1},
%\end{equation} 
%and similarly
%\begin{equation}\label{eq6}
%b=c^*e^{2i \varphi_1}
%\end{equation} 
%Then type-(5) spinors can be settled in a more representative way by\footnote{If $a \neq -d$ or $b \neq -c$, otherwise would be type-(2).}
%
%\begin{eqnarray}
%&\psi_{_{(5)}}=\left(\begin{tabular}{c}
%$-d^*e^{2i \varphi_1}$\\
%$c^*e^{2i \varphi_1}$\\
%$c$\\
%$d$
%\end{tabular}\right).\label{eq8}
%\end{eqnarray}
Writing it as $\psi_{_{(5)}}=\left(\eta, \xi \right)^T$, is easy to see that
 $\eta= -i\sigma_2\xi^* e^{2i\varphi_1}=\sigma_2\xi^* e^{i(2\varphi_1-\frac{\pi}{2})}$, and defining $\varphi\equiv 2\varphi_1-\frac{\pi}{2}$ we derive the eigenspinors of charge conjugation operator general form $\psi_{_{(5)}}=\left(e^{i\varphi}\sigma_2\xi^*, \xi \right)^T$, where
\begin{align}\label{conj}
\CCC\psi_{_{(5)}}=\zeta \psi_{_{(5)}},\quad{\rm  with}\quad \CCC=%
\begin{pmatrix}
\OO & \sigma_2 \\
-\sigma_2 & \OO  \nonumber
\end{pmatrix} \KKKK
\quad {\rm  and}\quad \zeta  =e^{-i\varphi}\,. 
\end{align}
The operator $\KKKK$ is responsible for the complex conjugation on spinors appearing on the right. 
Hence all type-(5) spinor field is eigenspinor of the charge conjugation operator with eigenvalues in the sphere $S^1$. 

\subsection*{Type-(4)}
%Up to my knowledge, until the year 2013 type-(4) spinors had never appeared in physics, but it was changed by **Cite, where was showed that in ESK gravitational theories type-(4) spinor raises as solution of the...**(um pouco mais sobre ESK) equations. 

For the type-(4) class we have just $\textbf{K}\neq 0$, $\textbf{S}\neq 0$ and $\sigma = 0 =\omega$. At first sight it could seen not so useful, but Lounesto showed that there are no more classes, besides all being disjoint. With this in mind we see that if  $a=0=b$ or $c=0=d$ we would have $\textbf{S}=0$, and thus it would correspond to a type-(6) spinor field. Subsequently it remains only  $c=0=b,\; a=0=d$, or all components do not equal zero.  By the condition $\textbf{K}\neq 0$, 
\begin{itemize}
\item If $c=0=b$ then $K_1 = K_2=0$ and $K_0=K_3$, thus $K_0 \neq 0 \Rightarrow ||a||^2 \neq ||d||^2$.

\item If $a=0=d$ we have a similar situation, leading to $K_0 \neq 0  \Rightarrow ||b||^2 \neq ||c||^2$.

\item If all the spinor field components are non zero $K_1 \neq 0 \neq K_2 \Rightarrow ||b||^2 \neq  ||c||^2$

\end{itemize}
In the last case $||b||^2 = ||c||^2$ implies $\textbf{K}=0$, yielding  a type-(5) spinor field. Moreover, it is straightforward to show that $||b||^2 \neq ||c||^2 \Leftrightarrow  ||a||^2 \neq ||d||^2$. Hence the possible type-(4) spinor fields are 
given by the most general formul\ae
\begin{align}
\psi_{_{(4)}}=\left(a,0,0,d\right)^T\; ;  \qquad \psi_{_{(4)}}=\left(0,b,c,0\right)^T\; ;  \qquad  \psi_{_{(4)}}=\left(-\frac{bcd^*}{||c||^2},b,c,d\right)^T.
\end{align}

\subsection*{Regular Spinor Fields}

For general regular spinor fields the only equation we have is $ a c^{*}+b d^{*} \neq 0$, hence we can extract just the following information

\begin{itemize}
\item If just $a$ and $c$ or just $b$ and $d$ are nonzero, the spinor field is singular.

\item If just one component is zero the spinor field  is singular.

\item If no one is zero and $a \neq -\frac{bcd^*}{||c||^2}$, the spinor field is singular.

\end{itemize}

\subsection*{Characterization table}

The  table above  summarizes all the results of this section, being  useful by giving a direct and easy way to decide whether a spinor field is regular, singular,  and what class it belongs to  (here ``n$^o$ of nz.c'' means ``number of non zero components'').

%\begin{landscape}
 \begin{tiny}
 \begin{table}[htbp]
 \title{Spinors Characterization Table}
 %\begin{landscape}
 \centering
 \setlength\parskip{0.2cm}
 \centering
 \begin{tabular}{c||c|c|c|c}
 \hline
 \hline
 \multicolumn{1}{c}{n$^o$ of nz.c}&%\backslashbox { n$^o$ de comp. não nulas }{ Tipo }} &
 \multicolumn{1}{c|}{1} & \multicolumn{1}{c|}{2} &
 \multicolumn{1}{c|}{3} & \multicolumn{1}{c}{4} \\
 \hline
 \hline
Classes &&&&\\
{Type-(4)} & -- & \begin{tabular}{c}

$\begin{pmatrix}
a\\
0\\
0\\
d
\end{pmatrix},\begin{pmatrix}
0\\
b\\
c\\
0
\end{pmatrix}$ \\ 

$\begin{array}{l}
 ||a||^2 \neq ||d||^2\\
  ||b||^2 \neq ||c||^2
\end{array}$ \\ 
\hline \\
\end{tabular}   & -- &\begin{tabular}{c}

 $\begin{pmatrix}
-\frac{bcd^*}{||c||^2}\\
b\\
c\\
d
\end{pmatrix}$\\ 
$ ||b||^2 \neq ||c||^2$\\ 
\hline 
\end{tabular} 
 \\
{Type-(5)} & -- &
\begin{tabular}{c}

$\begin{pmatrix}
a\\
0\\
0\\
d
\end{pmatrix},\begin{pmatrix}
0\\
b\\
c\\
0
\end{pmatrix}$ \\ 

$\begin{array}{l}
 ||a||^2 = ||d||^2\\
  ||b||^2 = ||c||^2
\end{array}$  \\ 
\hline \\
\end{tabular} 
 &-- & $\begin{pmatrix}
-d^*e^{2i \varphi_1}\\
c^*e^{2i \varphi_1}\\
c\\
d
\end{pmatrix}$  \\
{Type-(6)} &\; Arbitrary \;
%$\begin{pmatrix}
%a\\
%0\\
%0\\
%0
%\end{pmatrix},\begin{pmatrix}
%0\\
%b\\
%0\\
%0
%\end{pmatrix},\begin{pmatrix}
%0\\
%0\\
%c\\
%0
%\end{pmatrix},\begin{pmatrix}
%0\\
%0\\
%0\\
%d
%\end{pmatrix}$ 
& \begin{tabular}{c}
 
$\begin{pmatrix}
a\\
b\\
0\\
0
\end{pmatrix},\begin{pmatrix}
0\\
0\\
c\\
d
\end{pmatrix}$ \\ 
\hline 
\end{tabular} 
&-- & --\\
{Regular} & -- & $\begin{pmatrix}
a\\
0\\
c\\
0
\end{pmatrix},\begin{pmatrix}
0\\
b\\
0\\
d
\end{pmatrix}$ 
&\; Arbitrary \; 
%$\begin{pmatrix}
%a\\
%b\\
%c\\
%0
%\end{pmatrix},\begin{pmatrix}
%a\\
%b\\
%0\\
%d
%\end{pmatrix},\begin{pmatrix}
%a\\
%0\\
%c\\
%d
%\end{pmatrix},\begin{pmatrix}
%0\\
%b\\
%c\\
%d
%\end{pmatrix}$  
& \begin{tabular}{c}
$\begin{pmatrix}
a\\
b\\
c\\
d
\end{pmatrix}$ \\ 
$a \neq -\frac{bcd^*}{||c||^2}$
\end{tabular}    \\
 \hline
 \hline
% \caption{Caracterização de espinores}
 %\multicolumn{5}{l}{Fonte: Elaboração própria das autoras} \\
 \end{tabular}\caption{Singular Spinor Fields}
 \label{tab:codigos}
% \end{landscape}
 \end{table}
 \end{tiny}

\section{Final remarks}
%\label{aadkp}

In this work we have showed that the constraint equation of the Lounesto classification can be used to completely characterize singular spinor fields, besides giving information about regular spinor fields. The characterization is listed in the above table  and avoid calculations in the spinor fields classification. Moreover, the information  in the characterization table is the first step to  the classification of all possible spinor fields dynamics. Due to it, our first step here is to provide 
a complete characterization of all classes of singular spinor fields, in particular to pave the 
road to a complete characterization of all the possible dynamics for singular spinor fields in 
Lounesto classification. Moreover, once the general form of singular spinor fields is known, 
it is easier to devise what kind of type-(4) and type-(6) spinor fields, besides the well known Elko spinor fields, have mass dimension one. However it is out of the scope of this paper.
 
\section*{Acknowledgments}
 R. T. Cavalcanti is grateful to CAPES for financial support, and to Profs. R. da Rocha and J. M. Hoff da Silva for fruitful discussions.

\end{document}